%% file: main.tex
\documentclass[10pt, journal, twocolumn]{IEEEtran}

\usepackage{epsfig}
\usepackage{amsmath,amsthm,amssymb,amsfonts}
\usepackage{bm}

\usepackage{color}

\usepackage[noend]{algpseudocode}
\usepackage{algorithmicx,algorithm}
\usepackage{url}
\usepackage{hyperref}  

\begin{document}\sloppy

\def\x{{\mathbf x}}
\def\L{{\cal L}}

\title{Deep Reinforcement Learning for Automated Stock Trading: An Ensemble Strategy}
%
\author{Hongyang Yang$^{1}$, 
        Xiao-Yang Liu$^{2}$, 
        Shan Zhong$^{2}$, 
        and Anwar Walid$^3$\\
$^1$AI4Finance Foundation\thanks{The AI4Finance Foundation (https://ai4finance.org) is a U.S.-registered 501(c)(3) nonprofit public charity focused on promoting open scientific research in financial AI, building open-source infrastructure, and supporting a global community of researchers through shared datasets, benchmarks, and educational programs.}\\
$^2$Dept. of Electrical Engineering, Columbia University\\
$^3$Mathematics of Systems Research Department, Nokia-Bell Labs\\
Email: contact@ai4finance.org      
}

\maketitle

\begin{abstract}

Stock trading strategies play a critical role in investment. However, it is challenging to design a profitable strategy in a complex and dynamic stock market. In this paper, we propose an ensemble strategy that employs deep reinforcement schemes to learn a stock trading strategy by maximizing investment return. We train a deep reinforcement learning agent and obtain an ensemble trading strategy using three actor-critic based algorithms: Proximal Policy Optimization (PPO), Advantage Actor Critic (A2C), and Deep Deterministic Policy Gradient (DDPG). The ensemble strategy inherits and integrates the best features of the three algorithms, thereby robustly adjusting to different market situations. In order to avoid the large memory consumption in training networks with continuous action space, we employ a load-on-demand technique for processing very large data. We test our algorithms on the 30 Dow Jones stocks that have adequate liquidity. The performance of the trading agent with different reinforcement learning algorithms is evaluated and compared with both the Dow Jones Industrial Average index and the traditional min-variance portfolio allocation strategy. The proposed deep ensemble strategy is shown to outperform the three individual algorithms and two baselines in terms of the risk-adjusted return measured by the Sharpe ratio. This work is fully open-sourced at \href{https://github.com/AI4Finance-Foundation/Deep-Reinforcement-Learning-for-Automated-Stock-Trading-Ensemble-Strategy-ICAIF-2020}{GitHub}.

\end{abstract}


\input{sections/Introduction.tex}

\input{sections/RelatedWorks.tex}
\input{sections/ProblemDescription.tex}
\input{sections/Environment.tex}

\input{sections/Algorithms.tex}
\input{sections/Performance.tex}
\input{sections/Conclusion.tex}

\bibliographystyle{IEEEbib}
\bibliography{references}

\end{document}

%% file: sections/Introduction.tex
\section{Introduction}

Profitable automated stock trading strategy is vital to investment companies and hedge funds. It is applied to optimize capital allocation and maximize investment performance, such as expected return. Return maximization can be based on the estimates of potential return and risk. However, it is challenging for analysts to consider all relevant factors in a complex and dynamic stock market \cite{fuzzy, online, intelligence}.

\begin{figure}[t]
\centering
\includegraphics[scale=0.3]{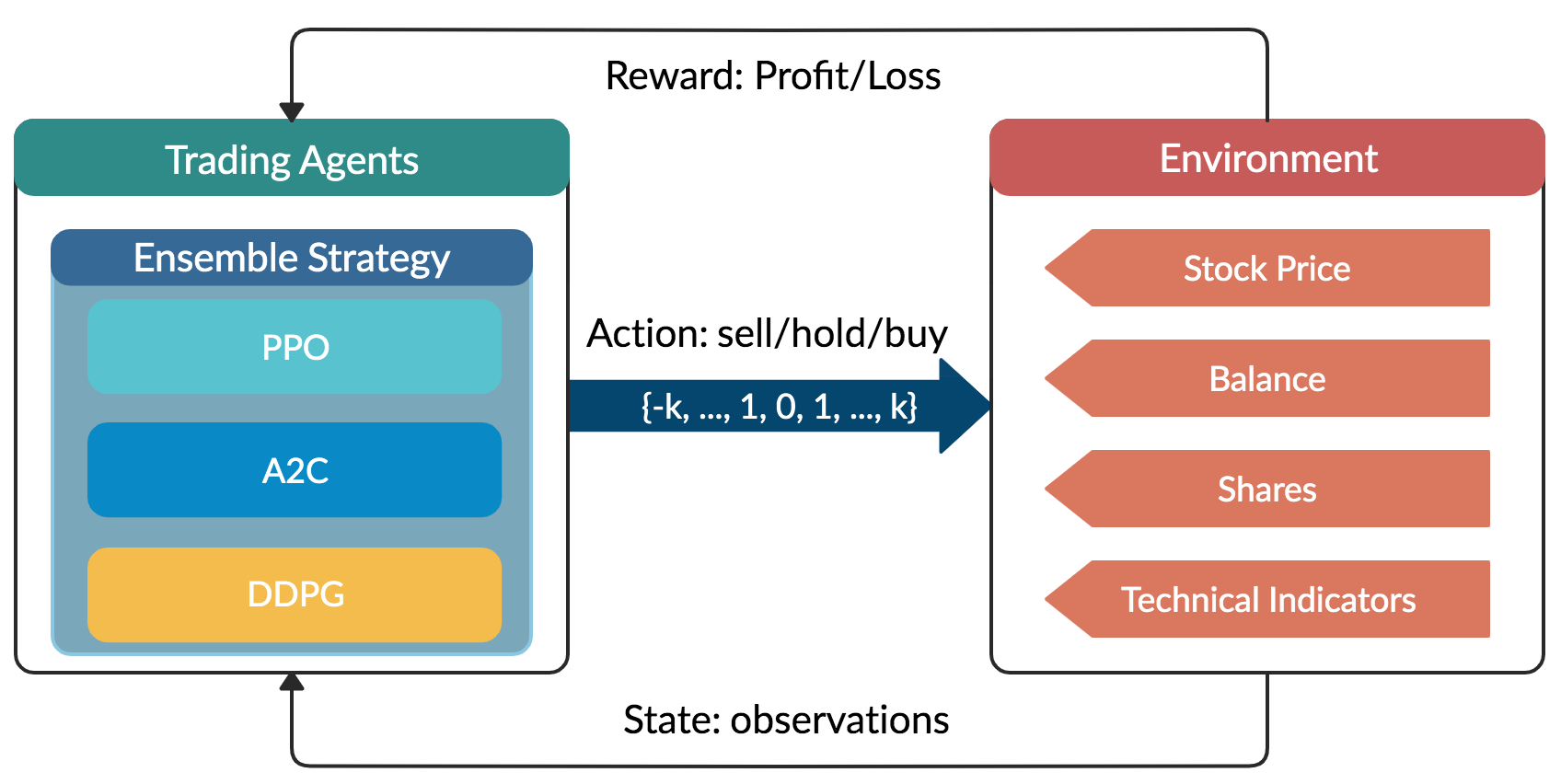}
\caption{Overview of reinforcement learning-based stock trading strategy.}
\label{trading}
\end{figure}

Existing works are not satisfactory. A traditional approach that employed two steps was described in \cite{selection}. First, the expected stock return and the covariance matrix of stock prices are computed. Then, the best portfolio allocation strategy can be obtained by either maximizing the return for a given risk ratio or minimizing the risk for a pre-specified return. This approach, however, is complex and costly to implement since the portfolio managers may want to revise the decisions at each time step, and take other factors into account, such as transaction cost. Another approach for stock trading is to model it as a Markov Decision Process (MDP) and use dynamic programming to derive the optimal strategy \cite{DP,Testing,Optimum, Enhance}. However, the scalability of this model is limited due to the large state spaces when dealing with the stock market. 

In recent years, machine learning and deep learning algorithms have been widely applied to build prediction and classification models for the financial market. Fundamentals data (earnings report) and alternative data (market news, academic graph data, credit card transactions, and GPS traffic, etc.) are combined with machine learning algorithms to extract new investment alphas or predict a company's future performance \cite{yang_2018, scholar_2019, news_2010, Qian}. Thus, a predictive alpha signal is generated to perform stock selection. However, these approaches are only focused on picking high performance stocks rather than allocating trade positions or shares between the selected stocks. In other words, the machine learning models are not trained to model positions.

In this paper, we propose a novel ensemble strategy that combines three deep reinforcement learning algorithms and finds the optimal trading strategy in a complex and dynamic stock market. The three actor-critic algorithms \cite{actor} are Proximal Policy Optimization (PPO) \cite{PPO_2017, Liang2018AdversarialDR}, Advantage Actor Critic (A2C) \cite{a3c_2016, zhang_2019}, and Deep Deterministic Policy Gradient (DDPG) \cite{DDPG, Liang2018AdversarialDR, Xiong2018PracticalDR}. 
Our deep reinforcement learning approach is described in Figure \ref{trading}.  
By applying the ensemble strategy, we make the trading strategy more robust and reliable. Our strategy can adjust to different market situations and maximize return subject to risk constraint. 
First, we build an environment and define action space, state space, and reward function. Second, we train the three algorithms that take actions in the environment. Third, we ensemble the three agents together using the Sharpe ratio that measures the risk-adjusted return. The effectiveness of the ensemble strategy is verified by its higher Sharpe ratio than both the min-variance portfolio allocation strategy and the Dow Jones Industrial Average \footnote{The Dow Jones Industrial Average is a stock market index that shows how 30 large, publicly owned companies based in the United States have traded during a standard trading session in the stock market.} (DJIA).

The remainder of this paper is organized as follows. Section 2 introduces related works. Section 3 provides a description of our stock trading problem. In Section 4, we set up our stock trading environment. In Section 5, we drive and specify the three actor-critic based algorithms and our ensemble strategy. Section 6 describes the stock data preprocessing and our experimental setup, and presents the performance evaluation of the proposed ensemble strategy. We conclude this paper in Section 7.

%% file: sections/RelatedWorks.tex
\section{Related Works}

Recent applications of deep reinforcement learning in financial markets consider discrete or continuous state and action spaces, and employ one of these learning approaches: critic-only approach, actor-only approach, or actor-critic approach \cite{RL_survey}. Learning models with continuous action space provide finer control capabilities than those with discrete action space.

The critic-only learning approach, which is the most common, solves a discrete action space problem using, for example, Deep Q-learning (DQN) and its improvements, and trains an agent on a single stock or asset \cite{DRL_automate, dang_RL, jeong_2018}. 
The idea of the critic-only approach is to use a Q-value function to learn the optimal action-selection policy that maximizes the expected future reward given the current state. Instead of calculating a state-action value table, DQN minimizes the error between estimated Q-value and target Q-value over a transition, and uses a neural network to perform function approximation. The major limitation of the critic-only approach is that
it only works with discrete and finite state and action spaces, which is not practical for a large portfolio of stocks, since the prices are of course continuous.

The actor-only approach has been used in \cite{moody_2001, deng_2016, jiang_2017}. The idea here is that the agent directly learns the optimal policy itself. Instead of having a neural network to learn the Q-value, the neural network learns the policy. The policy is a probability distribution that is essentially a strategy for a given state, namely the likelihood to take an allowed action. Recurrent reinforcement learning is introduced to avoid the curse of dimensionality and improves trading efficiency in \cite{moody_2001}. The actor-only approach can handle the continuous action space environments. 

The actor-critic approach has been recently applied in finance \cite{fuzzy_2010, li_a2c_2018,zhang_2019, Xiong2018PracticalDR}. The idea is to simultaneously update the actor network that represents the policy, and the critic network that represents the value function. The critic estimates the value function, while the actor updates the policy probability distribution guided by the critic with policy gradients. Over time, the actor learns to take better actions and the critic gets better at evaluating those actions. The actor-critic approach has proven to be able to learn and adapt to large and complex environments, and has been used to play popular video games, such as Doom \cite{Wu2017TrainingAF}. Thus, the actor-critic approach is promising in trading with a large stock portfolio.


%% file: sections/ProblemDescription.tex
\section{Problem Description}

We model stock trading as a Markov Decision Process (MDP), and formulate our trading objective as a maximization of expected return \cite{expected_return}. 
\subsection{MDP Model for Stock Trading}

\begin{figure}[t]
\centering
\includegraphics[scale = 0.35]{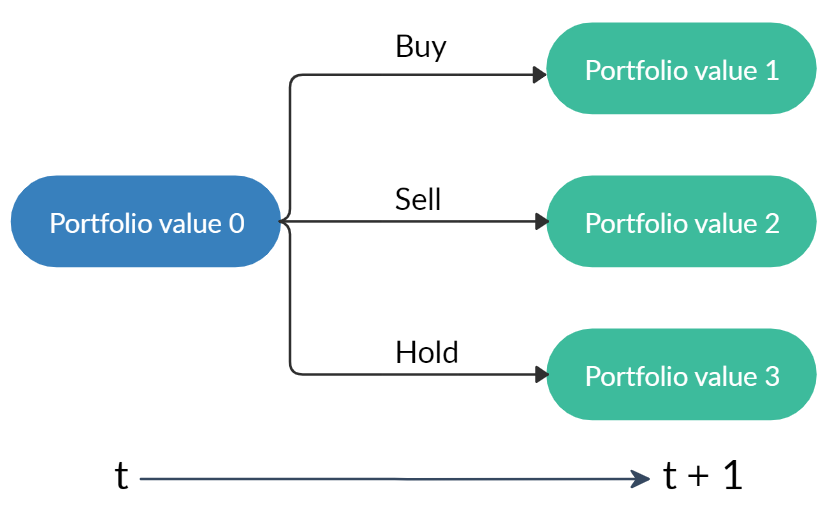}
\caption{A starting portfolio value with three actions result in three possible portfolios. Note that "hold" may lead to different portfolio values due to the changing stock prices.}
\label{iteration0}
\end{figure}

To model the stochastic nature of the dynamic stock market, we employ a Markov Decision Process (MDP) as follows:
\begin{itemize}
\item State $\bm{s}= [\bm{p},\bm{h},b]$: a vector that includes stock prices $\bm{p}\in \mathbb{R}_+^D$, the stock shares $\bm{h}\in \mathbb{Z}_+^D$, and the remaining balance $b\in \mathbb{R}_+$, where $D$ denotes the number of stocks and $\mathbb{Z}_+$ denotes non-negative integers. 
\item Action $\bm{a}$: a vector of actions over $D$ stocks. The allowed actions on each stock include \textit{selling}, \textit{buying}, or \textit{holding}, which result in decreasing, increasing, and no change of the stock shares $\bm{h}$, respectively. 
\item Reward $r(s,a,s')$: 
the direct reward of taking action $a$ at state $s$ and arriving at the new state $s'$. 
\item Policy $\pi(s)$: the trading strategy at state $s$, which is the probability distribution of actions at state $s$.
\item Q-value $Q_\pi(s,a)$: the expected reward of taking action $a$ at state $s$ following policy $\pi$.
\end{itemize}

The state transition of a stock trading process is shown in Figure \ref{iteration0}. At each state, one of three possible actions is taken on stock $d~(d=1, ...,D)$ in the portfolio.
\begin{itemize}
    \item Selling $\bm{k}[d]\in [1,\bm{h}[d]]$ shares results in $\bm{h_{t+1}}[d]=\bm{h_t}[d]-\bm{k}[d]$, where $\bm{k}[d] \in \mathbb{Z}_+$ and $d = 1,...,D$.
    \item Holding, $\bm{h_{t+1}}[d] = \bm{h_t}[d]$.
    \item Buying $\bm{k}[d]$ shares results in $\bm{h_{t+1}}[d]=\bm{h_t}[d]+\bm{k}[d]$.
\end{itemize}
At time $t$ an action is taken and the stock prices update at $t$+1, accordingly the portfolio values may change from "portfolio value 0" to "portfolio value 1", "portfolio value 2", or "portfolio value 3", respectively, as illustrated in Figure \ref{iteration0}. Note that the portfolio value is $\bm{p^T} \bm{h} + b$.

\subsection{Incorporating Stock Trading Constraints}
The following assumption and constraints reflect concerns for practice: transaction costs, market liquidity, risk-aversion, etc.

\begin{itemize}
\item Market liquidity: the orders can be rapidly executed at the close price. We assume that stock market will not be affected by our reinforcement trading agent.
\item Nonnegative balance $b\geq 0$: the allowed actions should not result in a negative balance. Based on the action at time $t$, the stocks are divided into sets for sell $\mathcal{S}$, buying $\mathcal{B}$, and holding $\mathcal{H}$, where $\mathcal{S}\cup\mathcal{B}\cup\mathcal{H} = \{1,\cdots,D\}$ and they are nonoverlapping.
Let $\bm{p}_t^B =[p_t^i: i \in \mathcal{B}]$ and $\bm{k}_t^B =[k_t^i: i \in \mathcal{B}]$ be the vectors of price and number of buying shares for the stocks in the buying set.
We can similarly define $\bm{p}_t^{S}$ and $\bm{k}_t^{S}$ for the selling stocks, and $\bm{p}_t^H$ and $\bm{k}_t^H$ for the holding stocks.
Hence, the constraint for non-negative balance can be expressed as
\begin{equation} 
b_{t+1} =b_t + (\bm{p}_t^S)^T \bm{k}_{t}^S 
-(\bm{p}_t^B)^T \bm{k}_{t}^B \geq 0.
\label{balance}
\end{equation} 
\item Transaction cost: transaction costs are incurred for each trade. There are many types of transaction costs such as exchange fees, execution fees, and SEC fees. Different brokers have different commission fees. Despite these variations in fees, we assume our transaction costs to be $0.1\%$ of the value of each trade (either buy or sell) as in \cite{yang_2018}:
\begin{equation}  
c_{t} = \bm{p}^T \bm{k}_{t} \times 0.1\%.
\label{cost}
\end{equation} 


\item Risk-aversion for market crash: there are sudden events that may cause stock market crash, such as wars, collapse of stock market bubbles, sovereign debt default, and financial crisis. To control the risk in a worst-case scenario like 2008 global financial crisis, we employ the financial turbulence index $turbulence_t$ that measures extreme asset price movements \cite{turbulence}:
\begin{equation}  
turbulence_t = \left(\bm{y_t} - \bm{\mu}\right)\bm{\Sigma^{-1}}(\bm{y_t}-\bm{\mu})' \in \mathbb{R},
\label{turb}
\end{equation} 
where $\bm{y_t} \in \mathbb{R}^D$ denotes the stock returns for current period t, $\bm{\mu} \in \mathbb{R}^D$ denotes the average of historical returns, and $\bm{\Sigma} \in \mathbb{R}^{D \times D}$ denotes the covariance of historical returns. When $turbulence_t$ is higher than a threshold, which indicates extreme market conditions, we simply halt buying and the trading agent sells all shares. We resume trading once the turbulence index returns under the threshold.
\end{itemize}

\subsection{Return Maximization as Trading Goal}


We define our reward function as the change of the portfolio value when action $a$ is taken at state $s$ and arriving at new state $s'$. The goal is to design a trading strategy that maximizes the change of the portfolio value:
\begin{align}
 r(s_t,a_t,s_{t+1}) =  (b_{t+1} + \bm{p_{t+1}}^T \bm{h_{t+1}}) - (b_{t} + \bm{p_{t}}^T \bm{h_{t}}) - c_{t},
 \label{reward1}
\end{align} 
where the first and second terms denote the portfolio value at $t+1$ and $t$, respectively. To further decompose the return, we define
the transition of the shares $\bm{h_{t}}$ is defined as
\begin{align}
 \bm{h}_{t+1} = \bm{h}_{t} - \bm{k}_{t}^S + \bm{k}_{t}^B, 
\end{align} 
and the transition of the balance $b_{t}$ is defined in (\ref{balance}).
Then (\ref{reward1}) can be rewritten as
\begin{align}
 r(s_t,a_t,s_{t+1}) =  r_H - r_S + r_B  - c_{t},
\label{reward2}
\end{align} where 
\begin{equation}  
\begin{split}
r_H = (\bm{p}_{t+1}^H-\bm{p}_{t}^H)^T \bm{h}_t^H,
\end{split}
\end{equation} 
\begin{align}
 r_S = (\bm{p}_{t+1}^S-\bm{p}_{t}^S)^T \bm{h}_t^S,
 \label{selling}
\end{align} 
\begin{align}
 r_B = (\bm{p}_{t+1}^B-\bm{p}_{t}^B)^T \bm{h}_t^B,
\end{align} 
where $ r_H$, $ r_S$, and $r_B$ denote the change of the portfolio value comes from holding, selling, and buying shares moving from time $t$ to $t+1$, respectively. Equation (\ref{reward2}) indicates that we need to maximize the positive change of the portfolio value by buying and holding the stocks whose price will increase at next time step and minimize the negative change of the portfolio value by selling the stocks whose price will decrease at next time step. 

Turbulence index $turbulence_t$ is incorporated with the reward function to address our risk-aversion for market crash. When the index in (\ref{turb}) goes above a threshold, Equation (\ref{selling}) becomes
\begin{align}
 r_{sell} = (\bm{p_{t+1}}-\bm{p_{t}})^T \bm{k_{t}},
\end{align} which indicates that we want to minimize the negative change of the portfolio value by selling all held stocks, because all stock prices will fall.

The model is initialized as follows. $p_0$ is set to the stock prices at time $0$ and $b_0$ is the amount of initial fund. The $h$ and $Q_{\pi}(s,a)$ are 0, and $\pi(s)$ is uniformly distributed among all actions for each state. Then, $Q_{\pi}(s_{t},a_{t})$ is updated through interacting with the stock market environment. The optimal strategy is given by the Bellman Equation, such that the expected reward of taking action $a_t$ at state $s_t$ is the expectation of the summation of the direct reward $r(s_t,a_t,s_{t+1})$ and the future reward in the next state $s_{t+1}$. Let the future rewards be discounted by a factor of $0 < \gamma < 1$ for convergence purpose, then we have
\begin{equation}
    \label{decom}
    Q_\pi(s_{t},a_t) = \mathbb{E}_{s_{t+1}}[r(s_t,a_t,s_{t+1}) + \gamma \mathbb{E}_{a_{t+1} \sim \pi(s_{t+1})}[ Q_\pi (s_{t+1},a_{t+1})]].
\end{equation} 

The goal is to design a trading strategy that maximizes the positive cumulative change of the portfolio value $r(s_t,a_t,s_{t+1})$ in the dynamic environment, and we employ the deep reinforcement learning method to solve this problem.

%% file: sections/Environment.tex
\section{Stock Market Environment}

Before training a deep reinforcement trading agent, we carefully build the environment to simulate real world trading which allows the agent to perform interaction and learning. In practical trading, various information needs to be taken into account, for example the historical stock prices, current holding shares, technical indicators, etc.
Our trading agent needs to obtain such information through the environment, and take actions defined in the previous section. We employ OpenAI gym to implement our environment and train the agent \cite{openai_gym, openai_baselines,stable-baselines}.

\subsection{Environment for Multiple Stocks }
We use a continuous action space to model the trading of multiple stocks. We assume that our portfolio has 30 stocks in total.
\subsubsection{State Space}
We use a 181-dimensional vector consists of seven parts of information to represent the state space of multiple stocks trading environment: $[{b}_{t}, \bm{p}_t, \bm{h}_{t}, \bm{M}_t,
\bm{R_t},\bm{C_t},\bm{X_t}]$. Each component is defined as follows:
\begin{itemize} 
\item ${b}_{t}\in \mathbb{R}_+$: available balance at current time step $t$. 
\item $\bm{p}_{t}\in \mathbb{R}_+^{30}$: adjusted close price of each stock. 
\item $\bm{h}_{t}\in \mathbb{Z}_+^{30}$: shares owned of each stock. 
\item $\bm{M}_t \in \mathbb{R}^{30}$: Moving Average Convergence Divergence (MACD) is calculated using close price. MACD is one of the most commonly used momentum indicator that identifies moving averages \cite{macd_rsi}. 
\item $\bm{R}_t \in \mathbb{R}_+^{30}$: Relative Strength Index (RSI) is calculated using close price. RSI quantifies the extent of recent price changes. If price moves around the support line, it indicates the stock is oversold, and we can perform the buy action. If price moves around the resistance, it indicates the stock is overbought, and we can perform the selling action. \cite{macd_rsi}. 
\item $\bm{C_t}\in \mathbb{R}_+^{30}$: Commodity Channel Index (CCI) is calculated using high, low and close price. CCI compares current price to average price over a time window to indicate a buying or selling action \cite{cci}.
\item $\bm{X_t}\in \mathbb{R}^{30}$:  Average Directional Index (ADX) is calculated using high, low and close price. ADX identifies trend strength by quantifying the amount of price movement  \cite{adx}. 
\end{itemize}

\subsubsection{Action Space}
For a single stock, the action space is defined as $\{-k,...,-1, 0, 1, ..., k\}$, where $k$ and $-k$ presents the number of shares we can buy and sell, and $k \le h_{max}$ while $h_{max}$ is a predefined parameter that sets as the maximum amount of shares for each buying action. Therefore the size of the entire action space is $(2k+1)^{30}$. The action space is then normalized to $[-1, 1]$, since the RL algorithms A2C and PPO define the policy directly on a Gaussian distribution, which needs to be normalized and symmetric  \cite{stable-baselines}.

\subsection{Memory Management}
\begin{figure}[t]
\centering
\includegraphics[scale=0.36]{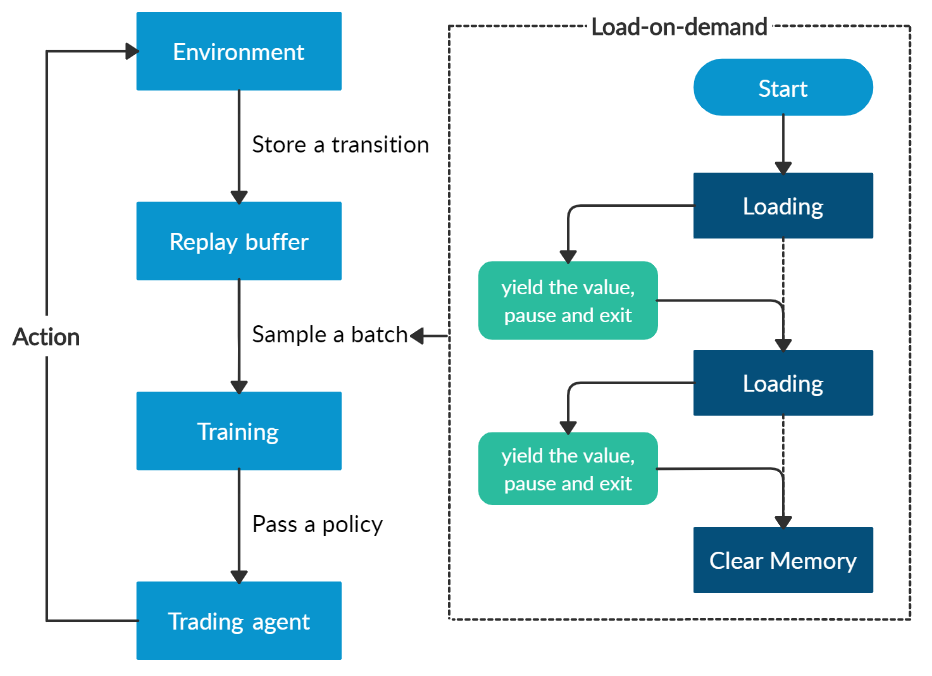}
\caption{Overview of the load-on-demand technique.}
\label{load}
\end{figure}

The memory consumption for training could grow exponentially with the number of stocks, data types, features of the state space, number of layers and neurons in the neural networks, and batch size. 
To tackle the problem of memory requirements, we employ a load-on-demand technique for efficient use of memory. As shown in Figure \ref{load}, the load-on-demand technique does not store all results in memory, rather, it generates them on demand. The memory is only used when the result is requested, hence the memory usage is reduced.

%% file: sections/Algorithms.tex
\section{Trading Agent based on Deep Reinforcement Learning}
We use three actor-critic based algorithms to implement our trading agent. The three algorithms are A2C, DDPG, and PPO, respectively. An ensemble strategy is proposed to combine the three agents together to build a robust trading strategy.

\subsection{Advantage Actor Critic (A2C)}
A2C \cite{a3c_2016} is a typical actor-critic algorithm and we use it a component in the ensemble strategy. A2C is introduced to improve the policy gradient updates. A2C utilizes an advantage function to reduce the variance of the policy gradient. Instead of only estimates the value function, the critic network estimates the advantage function. Thus, the evaluation of an action not only depends on how good the action is, but also considers how much better it can be. So that it reduces the high variance of the policy network and makes the model more robust. 

A2C uses copies of the same agent to update gradients with different data samples. Each agent works independently to interact with the same environment. In each iteration, after all agents finish calculating their gradients, A2C uses a coordinator to pass the average gradients over all the agents to a global network. So that the global network can update the actor and the critic network. The presence of a global network increases the diversity of training data. The synchronized gradient update is more cost-effective, faster and works better with large batch sizes. A2C is a great model for stock trading because of its stability.
 
The objective function for A2C is:
\begin{equation}
    \label{PG_objective}
     \nabla J_{\theta}(\theta) = \mathbb{E}[\sum_{t=1}^{T}\nabla_{\theta}\log \pi_{\theta}(a_{t}|s_{t})A(s_{t},a_{t}) ],
\end{equation} where $\pi_{\theta}(a_{t}|s_{t})$ is the policy network, $ A(s_{t},a_{t}) $ is the Advantage function can be written as:
\begin{equation}
 A(s_{t},a_{t}) = Q(s_{t},a_{t}) - V(s_{t}),
\end{equation} or 
\begin{equation}
 A(s_{t},a_{t}) = r(s_t,a_t,s_{t+1}) + \gamma V(s_{t+1}) - V(s_{t}).
\end{equation}




\subsection{Deep Deterministic Policy Gradient (DDPG)}

DDPG \cite{DDPG} is used to encourage maximum investment return. DDPG combines the frameworks of both Q-learning \cite{Sutton} and policy gradient \cite{policy_gradient_2000}, and uses neural networks as function approximators. In contrast with DQN that learns indirectly through Q-values tables and suffers the curse of dimensionality problem \cite{Management}, DDPG learns directly from the observations through policy gradient. It is proposed to deterministically map states to actions to better fit the continuous action space environment. 

At each time step, the DDPG agent performs an action $a_t$ at $s_t$, receives a reward $r_t$ and arrives at $s_{t+1}$. The transitions $(s_t,a_t,s_{t+1},r_t)$ are stored in the replay buffer $R$. A batch of $N$ transitions are drawn from $R$ and the Q-value $y_i$ is updated as: 
\begin{equation}
    y_i = r_i+\gamma Q' (s_{i+1}, \mu' (s_{i+1}|\theta^{\mu'},\theta^{Q'})), i=1,\cdots,N.
\end{equation} 
The critic network is then updated by minimizing the loss function $L(\theta^Q)$ which is the expected difference between outputs of the target critic network $Q'$ and the critic network $Q$, i.e,
\begin{align}
    L(\theta^Q) = \mathbb{E}_{s_t,a_t,r_t,s_{t+1} \sim \text{buffer}}[(y_i- Q(s_t,a_t|\theta^Q))^2].
\end{align}
DDPG is effective at handling continuous action space, and so it is appropriate for stock trading.




\subsection{Proximal Policy Optimization (PPO)}
We explore and use PPO as a component in the ensemble method. PPO \cite{PPO_2017} is introduced to control the policy gradient update and ensure that the new policy will not be too different from the previous one. PPO tries to simplify the objective of Trust Region Policy Optimization (TRPO) by introducing a clipping term to the objective function \cite{TRPO_2015,PPO_2017}. 

Let us assume the probability ratio between old and new policies is expressed as:
\begin{equation}
    r_t(\theta) = \frac{\pi_{\theta}(a_{t}|s_{t})}{\pi_{\theta_{old}}(a_{t}|s_{t})}.
\end{equation}
The clipped surrogate objective function of PPO is:
\begin{equation}
 \begin{split}
    J^{\text{CLIP}}(\theta) = 
    &\mathbb{\hat{E}}_{t}[\min(r_t(\theta)\hat{A}(s_{t},a_{t}),\\
    & \text{clip}(r_t(\theta), 1-\epsilon,1+\epsilon)\hat{A}(s_{t},a_{t}))], 
 \end{split}
\end{equation} where $r_t(\theta)\hat{A}(s_{t},a_{t})$ is the normal policy gradient objective, and $\hat{A}(s_{t},a_{t})$ is the estimated advantage function. The function $clip(r_t(\theta), 1-\epsilon, 1+\epsilon)$ clips the ratio $r_t(\theta)$ to be within $[1-\epsilon,1+\epsilon]$. The objective function of PPO takes the minimum of the clipped and normal objective. PPO discourages large policy change move outside of the clipped interval. Therefore, PPO improves the stability of the policy networks training by restricting the policy update at each training step. We select PPO for stock trading because it is stable, fast, and simpler to implement and tune.


\subsection{Ensemble Strategy}
Our purpose is to create a highly robust trading strategy. So we use an ensemble strategy to automatically select the best performing agent among PPO, A2C, and DDPG to trade based on the Sharpe ratio. The ensemble process is described as follows:

\textbf{Step 1}. We use a growing window of $n$ months to retrain our three agents concurrently. In this paper we retrain our three agents at every three months. 

\textbf{Step 2}. We validate all three agents by using a 3-month validation rolling window after training window to pick the best performing agent with the highest Sharpe ratio \cite{Sharpe}. The Sharpe ratio is calculated as:
\begin{equation}
    Sharpe~ratio = \frac{\bar{r}_{p}-r_f}{\sigma_p},
\end{equation} where $\bar{r}_{p}$ is the expected portfolio return, $r_f$ is the risk free rate, and $\sigma_p$ is the portfolio standard deviation. We also adjust risk-aversion by using turbulence index in our validation stage.

\textbf{Step 3}. After the best agent is picked, we use it to predict and trade for the next quarter.

The reason behind this choice is that each trading agent is sensitive to different type of trends. One agent performs well in a bullish trend but acts bad in a bearish trend. Another agent is more adjusted to a volatile market. The higher an agent's Sharpe ratio, the better its returns have been relative to the amount of investment risk it has taken. Therefore, we pick the trading agent that can maximize the returns adjusted to the increasing risk.

%% file: sections/Performance.tex
\section{Performance Evaluations}

In this section, we present the performance evaluation of our proposed scheme. We perform backtesting for the three individual agents and our ensemble strategy. The result in Table 2 demonstrates that our ensemble strategy achieves higher Sharpe ratio than the three agents, Dow Jones Industrial Average and the traditional min-variance portfolio allocation strategy.

Our codes are available on Github \footnote{Link: \url{https://github.com/AI4Finance-Foundation/Deep-Reinforcement-Learning-for-Automated-Stock-Trading-Ensemble-Strategy-ICAIF-2020}}.

\subsection{Stock Data Preprocessing}

We select the Dow Jones $30$ constituent stocks (at 01/01/2016) as our trading stock pool. Our backtestings use historical daily data from 01/01/2009 to 05/08/2020 for performance evaluation. The stock data can be downloaded from the Compustat database through the Wharton Research Data Services (WRDS) \cite{wrds}. 
Our dataset consists of two periods: in-sample period and out-of-sample period. In-sample period contains data for training and validation stages. Out-of-sample period contains data for trading stage. In the training stage, we train three agents using PPO, A2C, and DDPG, respectively. Then, a validation stage is then carried out for validating the 3 agents by Sharpe ratio, and adjusting key parameters, such as learning rate, number of episodes, etc. Finally, in the trading stage, we evaluate the profitability of each of the algorithms. 

\begin{figure}[bt]
\centering
\includegraphics[scale=0.45]{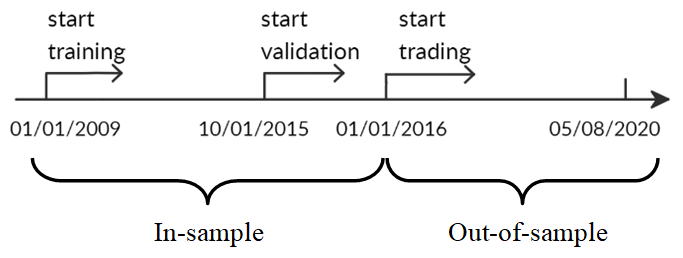}
\caption{Stock data splitting.}
\label{data}
\end{figure}

The whole dataset is split as shown in Figure \ref{data}. Data from 01/01/2009 to 09/30/2015 is used for training, and the data from 10/01/2015 to 12/31/2015 is used for validation and tuning of parameters. Finally, we test our agent's performance on trading data, which is the unseen out-of-sample data from 01/01/2016 to 05/08/2020. To better exploit the trading data, we continue training our agent while in the trading stage, since this will help the agent to better adapt to the market dynamics.

\subsection{Performance Comparisons}

\subsubsection{Agent Selection}
From Table 1, we can see that PPO has the best validation Sharpe ratio of 0.06 from 2015/10 to 2015/12, so we use PPO to trade for the next quarter from 2016/01 to 2016/03. DDPG has the best validation Sharpe ratio of 0.61 from 2016/01 to 2016/03, so we use DDPG to trade for the next quarter from 2016/04 to 2016/06. A2C has the best validation Sharpe ratio of -0.15 from 2020/01 to 2020/03, so we use A2C to trade for the next quarter from 2020/04 to 2020/05. Five metrics are used to evaluate our results: 
\begin{itemize}
  \item Cumulative return: is calculated by subtracting the portfolio's final value from its initial value, and then dividing by the initial value.
  \item Annualized return: is the geometric average amount of money earned by the agent each year over the time period.
  \item Annualized volatility: is the annualized standard deviation of portfolio return.
  \item Sharpe ratio: is calculated by subtracting the annualized risk free rate from the annualized return, and the dividing by the annualized volatility.
  \item Max drawdown: is the maximum percentage loss during the trading period.
\end{itemize}
\begin{table}[ht]
    \centering
    \caption{Sharpe Ratios over time.}
    \begin{tabular}{|c|c|c|c|c|}
    \hline
        Trading Quarter&PPO  & A2C &  DDPG &  Picked Model \\
         \hline
         2016/01-2016/03 & \textbf{0.06}  & 0.03 & 0.05 &PPO\\
        \hline
        2016/04-2016/06 & 0.31  & 0.53 & \textbf{0.61} &DDPG\\        
        \hline
        2016/07-2016/09 & -0.02  & 0.01 & \textbf{0.05} &DDPG\\        
         \hline
        2016/10-2016/12 & \textbf{0.11}  & 0.01 & 0.09 &PPO\\        
         \hline
        2017/01-2017/03 & \textbf{0.53}  & 0.44 & 0.13 &PPO\\        
         \hline
        2017/04-2017/06 & 0.29  & \textbf{0.44} & 0.12 &A2C\\        
         \hline
        2017/07-2017/09 & \textbf{0.4}  & 0.32 & 0.15 &PPO\\        
         \hline
        2017/10-2017/12 & -0.05  & -0.04 & \textbf{0.12} &DDPG\\        
         \hline
        2018/01-2018/03 & \textbf{0.71}  & 0.63 & 0.62 &PPO\\        
         \hline
        2018/04-2018/06 & -0.08  & -0.02 & \textbf{-0.01} &DDPG\\        
         \hline
        2018/07-2018/09 & -0.17  & \textbf{0.21} & -0.03 &A2C\\        
         \hline
        2018/10-2018/12 & 0.30  & \textbf{0.48} & 0.39 &A2C\\        
         \hline
        2019/01-2019/03 & -0.26  & -0.25 & \textbf{-0.18} &DDPG\\        
         \hline
        2019/04-2019/06 & \textbf{0.38}  & 0.29 & 0.25 &PPO\\        
         \hline
        2019/07-2019/09 & \textbf{0.53}  & 0.47 & 0.52 &PPO\\        
         \hline
        2019/10-2019/12 & -0.22  & \textbf{0.11} & -0.22 &A2C\\        
         \hline
        2020/01-2020/03 & -0.36  & \textbf{-0.13} & -0.22 &A2C\\        
         \hline
        2020/04-2020/05 & -0.42  & \textbf{-0.15} & -0.58 &A2C\\        
         \hline
    \end{tabular}
\end{table}
Cumulative return reflects returns at the end of trading stage. Annualized return is the return of the portfolio at the end of each year. Annualized volatility and max drawdown measure the robustness of a model. The Sharpe ratio is a widely used metric that combines the return and risk together.

\begin{figure*}
\centering
\includegraphics[width=1\textwidth]{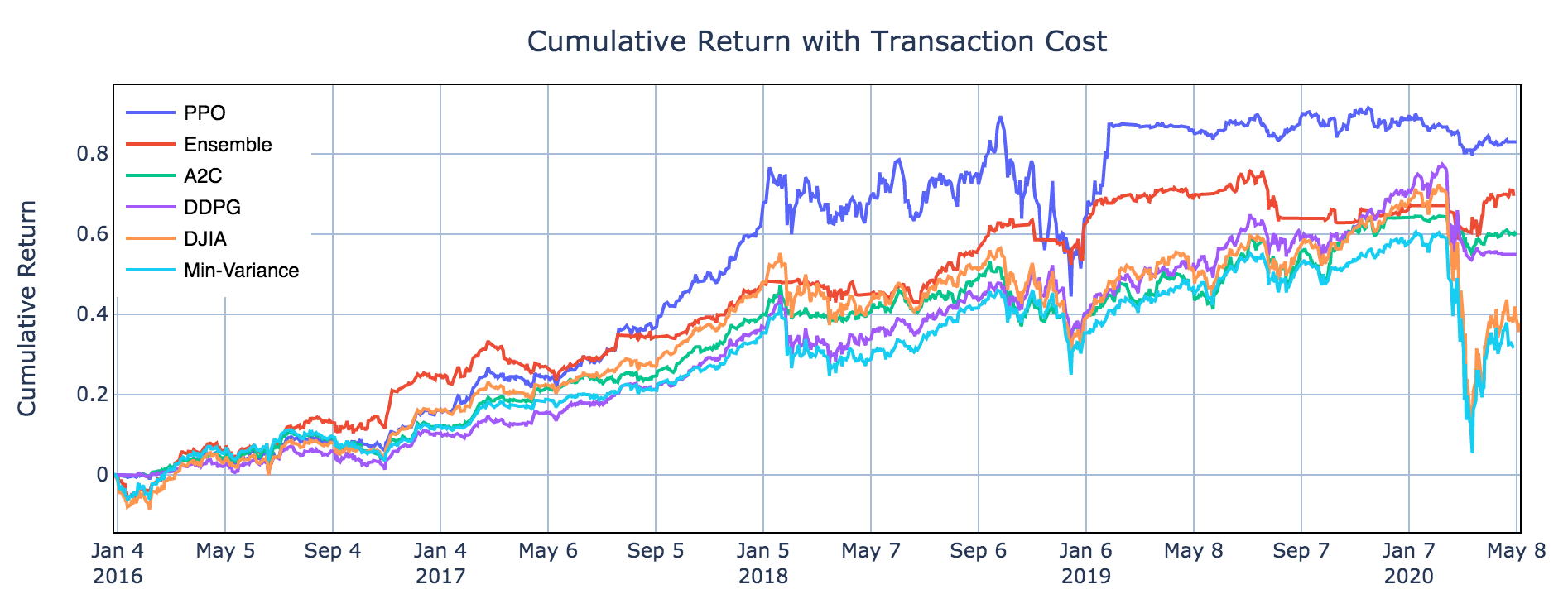}
\caption{Cumulative return curves of our ensemble strategy and three actor-critic based algorithms, the min-variance portfolio allocation strategy, and the Dow Jones Industrial Average. (Initial portfolio value $\$1,000,000$, from 2016/01/04 to 2020/05/08).}
\label{comparing}
\end{figure*}

\begin{table*}[htb]
\centering
\caption{Performance evaluation comparison.}
\begin{tabular}{|l|c|c|c|c|c|c|}
\hline
\textbf{(2016/01/04-2020/05/08)}         & \textbf{Ensemble (Ours)} & \textbf{PPO} & \textbf{A2C} & \textbf{DDPG} & \textbf{Min-Variance}& \textbf{DJIA} \\ \hline
\textbf{Cumulative Return}   & 70.4\%                   & 83.0\%                   & 60.0\%                         & 54.8\%                 & 31.7\%  & 38.6\%                  \\ \hline
\textbf{Annual Return}       & 13.0\%                     & 15.0\%                    & 11.4\%                          & 10.5\%                  & 6.5\%     & 7.8\%                \\ \hline
\textbf{Annual Volatility}   & 9.7\%                     & 13.6\%                    & 10.4\%                          & 12.3\%                    & 17.8\%  & 20.1\%                  \\ \hline
\textbf{Sharpe Ratio}        & 1.30                       & 1.10                      & 1.12                           & 0.87                     & 0.45   & 0.47                  \\ \hline
\textbf{Max Drawdown}        & -9.7\%                    & -23.7\%                   & -10.2\%                         & -14.8\%                  & -34.3\%   & -37.1\%               \\ \hline
\end{tabular}
\end{table*}

\begin{figure*}[ht]
\centering
\includegraphics[width=1\textwidth]{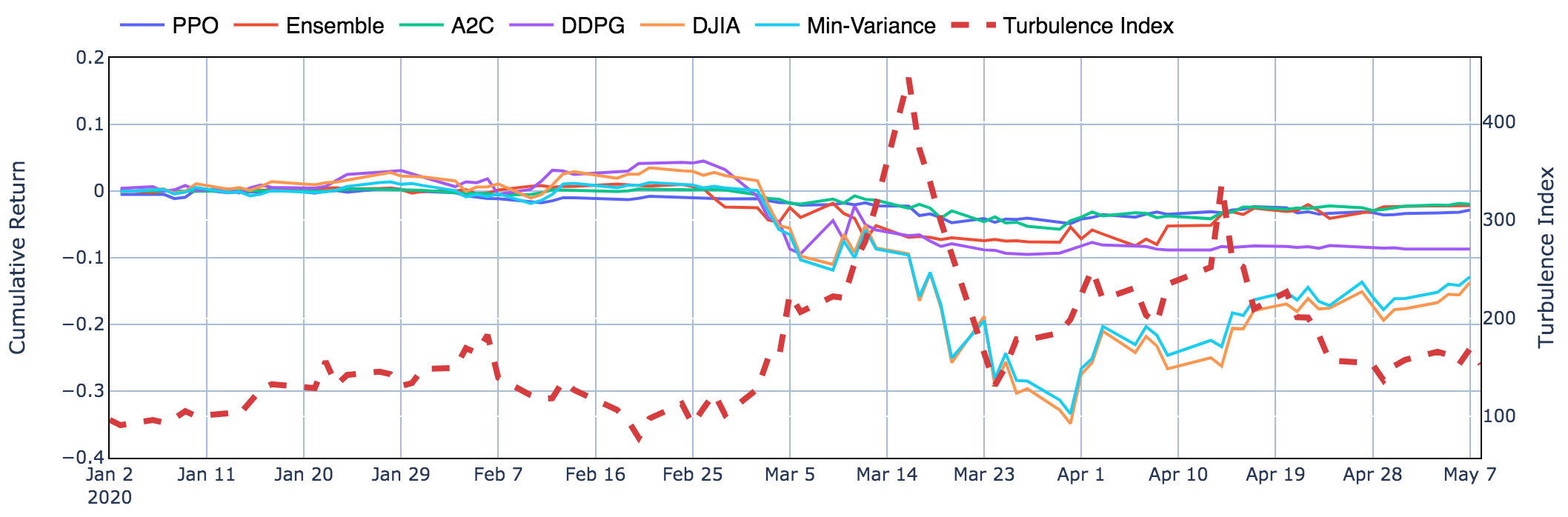}
\caption{Performance during the stock market crash in the first quarter of 2020.}
\label{trading_curb}
\end{figure*}

\subsubsection{Analysis of Agent Performance}
From both Table 2 and  Figure \ref{comparing}, we can observe that the A2C agent is more adaptive to risk. It has the lowest annual volatility $10.4\%$ and max drawdown $-10.2\%$ among the three agents. So A2C is good at handling a bearish market. PPO agent is good at following trend and acts well in generating more returns, it has the highest annual return $15.0\%$ and cumulative return $83.0\%$ among the three agents. So PPO is preferred when facing a bullish market. DDPG performs similar but not as good as PPO, it can be used as a complementary strategy to PPO in a bullish market. All three agents' performance outperform the two benchmarks, Dow Jones Industrial Average and min-variance portfolio allocation of DJIA, respectively.
\subsubsection{Performance under Market Crash}
In Figure \ref{trading_curb}, we can see that our ensemble strategy and the three agents perform well in the 2020 stock market crash event. When the turbulence index reaches a threshold, it indicates an extreme market situation. Then our agents will sell off all currently held shares and wait for the market to return to normal to resume trading. By incorporating the turbulence index, the agents are able to cut losses and successfully survive the stock market crash in March 2020. We can tune the turbulence index threshold lower for higher risk aversion.
\subsubsection{Benchmark Comparison}
Figure \ref{comparing} demonstrates that our ensemble strategy significantly outperforms the DJIA and the min-variance portfolio allocation \cite{yang_2018}. 
As can be seen from Table 2, the ensemble strategy achieves a Sharpe ratio $1.30$, which is much higher than the Sharpe ratio of $0.47$ for DJIA, and $0.45$ for the min-variance portfolio allocation. The annualized return of the ensemble strategy is also much higher, the annual volatility is much lower, indicating that the ensemble strategy beats both the DJIA and min-variance portfolio allocation in balancing risk and return. The ensemble strategy also outperforms A2C with a Sharpe ratio of 1.12, PPO with a Sharpe ratio of 1.10, and DDPG with a Sharpe ratio of 0.87, respectively. Therefore, our findings demonstrate that the proposed ensemble strategy can effectively develop a trading strategy that outperforms the three individual algorithms and the two baselines.


%% file: sections/Conclusion.tex
\section{Conclusion}

In this paper, we have explored the potential of using actor-critic based algorithms which are Proximal Policy Optimization (PPO), Advantage Actor Critic (A2C), and Deep Deterministic Policy Gradient (DDPG) agents to learn stock trading strategy. In order to adjust to different market situations, we use an ensemble strategy to automatically select the best performing agent to trade based on the Sharpe ratio. Results show that our ensemble strategy outperforms the three individual algorithms, the Dow Jones Industrial Average and min-variance portfolio allocation method in terms of Sharpe ratio by balancing risk and return under transaction costs. 

For future work, it will be interesting to explore more sophisticated model \cite{RNN}, solve empirical challenges \cite{DulacArnold2020AnEI}, deal with large-scale data \cite{Large} such as S\&P 500 constituent stocks. We can also explore more features for the state space such as adding advanced transaction cost and liquidity model \cite{bao_2019}, incorporating fundamental analysis indicators \cite{yang_2018}, natural language processing analysis of financial market news \cite{xinyi_2019}, and ESG features \cite{Qian} to our observations. We are interested in directly using Sharpe ratio as the reward function, but the agents need to observe a lot more historical data, the state space will increase exponentially.  
 